\documentclass[showpacs,nofootinbib,preprintnumbers,prd,superscriptaddress,twocolumn]{revtex4}

\usepackage{amsmath}
\usepackage{amsfonts}
\usepackage{amssymb}
\usepackage{graphicx}
\usepackage[titletoc]{appendix}
\usepackage{color}
\usepackage{hyperref}
\usepackage{cleveref}
\usepackage[rightcaption]{sidecap}
\usepackage{subfigure}
\usepackage{comment}
\usepackage{leftindex}

\usepackage{mathtools}

\usepackage{dcolumn}

\usepackage{array}
\usepackage{ctable}
\usepackage{multirow}
\usepackage{siunitx}
\usepackage{longtable}
\usepackage{tabularx}
\usepackage{booktabs}

\graphicspath{{Graphics/}}

\def\be{\begin{equation}}
\def\ee{\end{equation}}
\def\bea{\begin{eqnarray}}
\def\eea{\end{eqnarray}}

\definecolor{vividviolet}{rgb}{0.62, 0.0, 1.0}
\definecolor{amaranth}{rgb}{0.9, 0.17, 0.31}
\definecolor{palatinateblue}{rgb}{0.15, 0.23, 0.89}
\definecolor{brightpink}{rgb}{1.0, 0.0, 0.5}
\definecolor{cornflowerblue}{rgb}{0.39, 0.58, 0.93}
\definecolor{deepcarminepink}{rgb}{0.94, 0.19, 0.22}
\definecolor{radicalred}{rgb}{1.0, 0.21, 0.37}

\hypersetup{ linktoc=all,
    colorlinks, linkcolor={palatinateblue},
    citecolor={brightpink}, urlcolor={amaranth}
}

\begin{document}

\title{Production of ultralight dark matter from  inflationary spectator fields}

\author{Alessio Belfiglio}
\email{alessio.belfiglio@unicam.it}
\affiliation{School of Science and Technology, University of Camerino, Via Madonna delle Carceri, Camerino, 62032, Italy.}
\affiliation{Istituto Nazionale di Fisica Nucleare (INFN), Sezione di Perugia, Perugia, 06123, Italy.}

\author{Orlando Luongo}
\email{orlando.luongo@unicam.it}
\affiliation{School of Science and Technology, University of Camerino, Via Madonna delle Carceri, Camerino, 62032, Italy.}
\affiliation{Istituto Nazionale di Fisica Nucleare (INFN), Sezione di Perugia, Perugia, 06123, Italy.}
\affiliation{SUNY Polytechnic Institute, 13502 Utica, New York, USA.}
\affiliation{INAF - Osservatorio Astronomico di Brera, Milano, Italy.}
\affiliation{Al-Farabi Kazakh National University, Al-Farabi av. 71, 050040 Almaty, Kazakhstan.}

\begin{abstract}
    We investigate inflationary particle production associated with a spectator ultralight scalar field, which has been recently proposed as a plausible dark matter candidate. In this framework, we select the Starobinsky potential to drive the inflationary epoch, also discussing the case of a nonminimally coupled inflaton field fueled by a quartic symmetry-breaking potential. We focus on particle production arising from spacetime perturbations, which are induced by inflaton fluctuations during the quasi-de Sitter stage of inflation. In particular, we construct the first order Lagrangian describing interaction between inhomogeneities and the spectator field, quantifying superhorizon particle production during slow-roll. We then compare this mechanism with gravitational particle production associated with an instantaneous transition from inflation to the radiation dominated era. We show that the amount of particles obtained from perturbations is typically non-negligible and it is significantly enhanced on super-Hubble scales by the nonadiabatic inflationary expansion. Possible implications for primordial entanglement generation are also debated. 
\end{abstract}


\maketitle


\section{Introduction}\label{intro}

Dark matter (DM) is undoubtedly a key ingredient
to explain the cosmological large-scale dynamics and clustering \cite{dmrev}. Its nature, however, remains mysterious: we are still lacking any conclusive experiment able to unambiguously identify its properties, with slight agreement among the \emph{plethora} of theoretical proposals \cite{dm1}. 

Within this scenario, recent efforts in the search for weakly interacting massive particles from a few up to $100$ GeV have unfortunately proved unsuccessful \cite{dd1,dd2,dd3,dd4}. Thus, this lack of evidence has also revived the interest in \emph{ultralight} DM candidates such as axions \cite{ax0,ax1,ax2,ax3,ax4,ax5}, axion-like particles \cite{qax1,qax2,qax3}, ``fuzzy" DM models \cite{fuz1,fuz2,fuz3,udm1} and so on.

In this respect, several treatments have been proposed to explain the origin of DM particles. Among them, gravitational particle production (GPP)  represents a plausible and widely-investigated approach  \cite{gpp1,gpp2,gpp3,gpp4} as it creates particles directly from vacuum fluctuations and does not require any coupling between DM and generic quantum fields\footnote{Indeed the energy for particle creation is \emph{directly} obtained from the universe expansion, taking into account only the Einstein's field equations.}. For this reason, GPP of DM has been studied in various cosmological contexts, with particular interest on inflationary \cite{gpinfl1, gpinfl2,gpinfl3,gpinfl4,gpinfl5} and reheating \cite{gpreh1,gpreh2,gpreh3} phases. 

Gravitational production of ultralight particles has been recently discussed in Ref. \cite{boya}, focusing on the dynamics of a \emph{spectator scalar field} in the transition between inflation and the radiation dominated era. There, assuming an instantaneous transition, and thus neglecting the details of reheating, a significant particle production may take place for super-Hubble wavelength modes after inflation, if the field starts from the Bunch-Davies vacuum \cite{bunch1,bunch2,bunch3} and it is minimally coupled to spacetime curvature. Specifically, the matching conditions between the two epochs provide the Bogoliubov coefficients from which one obtains the number density of produced particles, compatible with a \emph{cold ultralight DM candidate}\footnote{GPP turns out to be more efficient during the nonadiabatic regime of inflationary expansion, modeled by an exact de Sitter phase. Adiabaticity is gradually recovered during the radiation era, so that a proper definition of particle states is again possible before matter-radiation equality.}.

This approach, however, neglects the slow-roll of the inflaton field and the dynamics of its quantum fluctuations, which represent the fundamental seeds for structure formation in our universe \cite{pert1,pert2, bau, rio}. In fact, inflationary fluctuations induce perturbations on the de Sitter dynamics of background, and the presence of inhomogeneities, by virtue of expansion, may result in the production of additional ``geometric" particles, due to purely gravitational effects \cite{fri, ces}. In Ref. \cite{giam}, it was argued that DM could be reinterpreted as geometric quasi-particles, in the attempt of solving the cosmological constant problem \cite{eone}, extending a mechanism of direct cancellation between fields \cite{etwo,ethree}.

Thus, at least two more points require additional investigations. First, how the nonminimal coupling acts on the inflationary dynamics and particle production, see e.g. Refs. \cite{efour,efive}, and, second, whether gravitational production furnishes stable particles, or more broadly, stable quasi-particles, since their dynamics at the end of the slow-roll regime can be altered by possible couplings of the inflaton to other quantum fields, as expected in the standard picture of reheating \cite{reh1,reh2,reh3}.

For the above reasons, we here aim to generalize inflationary geometric production to the case of a spectator scalar field. Assuming a quasi-de Sitter background evolution to properly account for the slow-roll of the inflaton, we show how spacetime perturbations generated by inflaton fluctuations couple to the energy-momentum tensor of a given spectator field, leading to particle creation during the slow-roll regime.  We focus in particular on the Starobinsky model of inflation \cite{staro1,staro2,staro3}, also discussing the case of a nonminimally coupled inflaton field driven by a quartic potential \cite{futa,faki}. We single out these paradigms since both the above models are among the best options to describe inflation, as certified by the Planck satellite measurements \cite{planck}.  

We observe that the amount of created particles depends on the mass of the spectator field and on the details of its coupling to the background. In particular, the presence of a small but still non-negligible coupling to the scalar curvature of spacetime is able to produce a significant amount of particles for super-Hubble modes\footnote{Geometric production is typically enhanced when dealing with large-field inflationary scenarios. In case of small-field inflation (e.g. hilltop models \cite{hill}), perturbations associated with inflaton fluctuations are necessarily smaller, thus resulting in  a much lower amount of produced particles (see, e.g., Ref. \cite{bel1}).}. Low-momentum enhancement of particle production is a peculiar trait of bosonic fields, and it is similarly found in unperturbed GPP scenarios. However, the presence of inhomogeneities allows for mode-mixing in particle production\footnote{On the other hand, unperturbed GPP only results in particle pairs with equal and opposite momenta, since the total momentum is necessarily conserved in homogeneous spacetimes \cite{gpp4}.}, with the Hubble radius emerging as the natural separation scale for modes during inflation. Consequently, motivated by these facts, we investigate particle production across the Hubble horizon, showing that a perturbative treatment is possible for super-Hubble modes that crossed the horizon well before the end of inflation. Since this approach was recently employed to study the entropy of cosmological perturbations \cite{ent1,ent2,ent3,ent4}, we accordingly work the inclusion of geometric and perturbative effects in primordial particle creation mechanisms out, showing that this is not only required for computing the correct abundance of DM candidates, but it may also shed further light on the quantum properties and entropy associated with the created particles. Physical consequences of our approach are promising, confirming that the nature of DM may arise from a spectator field, subdominant throughout the inflationary evolution.

The work is organized as follows. In Sec. \ref{sec2}, we discuss the features of the spectator DM field, computing the Bogoliubov coefficients related to GPP. In Sec. \ref{sec3}, we focus on the geometric contribution to GPP. In Sec. \ref{sezione4}, we analyze the main consequences of our results throughout inflationary stages, whereas in Sec. \ref{sezione5} we emphasize the main results inferred from our findings, highlighting possible quantum signatures, detectable at late-times. Conclusions and perspectives are reported in Sec. \ref{sezione6}.  \\


\section{Spectator field dynamics} \label{sec2}

We assume that, besides the inflaton, a subdominant field is present throughout the inflationary phase, with no interaction with the inflaton itself. Consequently, we exclusively consider gravitational interactions, disregarding potential couplings to other quantum fields both during inflation and the subsequent radiation era.

In this respect, we consider the Lagrangian density for the spectator, $\varphi$, with mass $m$, 
\be \label{spelag}
\mathcal{L}_S= \frac{1}{2} \left[ g^{\mu \nu} \varphi_{, \mu} \varphi_{,\nu}- \left( m^2+\xi_\varphi R \right) \varphi^2  \right],
\ee
where, as above stated, the only interaction is with the background, i.e., we include a nonminimal coupling between the curvature and $\varphi$, while the subscript ``S'' indicates the nature of $\varphi$. 

Further, in Eq. \eqref{spelag}, $g$ is the spacetime metric determinant, while $\xi_\varphi$ describes the coupling strength between the spectator field and the Ricci background scalar, $R$. 

To claim that $\varphi$ represents a spectator field, we require that its energy density is sufficiently small not to affect the inflationary dynamics. Accordingly, its evolution during the slow-roll phase is solely determined by the background potential driving inflation. To this end, we now describe the inflaton dynamics, then selecting the corresponding inflationary potentials in fulfillment of the most recent developments provided by the Planck satellite measurements \cite{planck}.


\subsection{Setting up the inflationary scenario} \label{sec2A}

During inflation, the background evolution is governed by the inflaton field, $\phi$, which we assume of scalar nature. The corresponding Lagrangian density reads 
\be \label{inflag}
\mathcal{L}_I= \frac{1}{2}  g^{\mu \nu} \phi_{, \mu} \phi_{,\nu}- V(\phi)  ,
\ee
where the potential $V(\phi)$ dominates over the other species and the subscript ``I'' denotes the inflationary epoch.

The inflationary standard paradigm predicts the seeds for gravitational small perturbations, induced by perturbing the inflaton through the standard ansatz \cite{pert1,pert2}
\be \label{infans}
\phi({\bf x},\tau)=\phi_0(\tau)+ \delta \phi ({\bf x},\tau),
\ee
where the homogeneous background term, $\phi_0$, is distinct from its corresponding quantum fluctuations, denoted by $\delta \phi$, depending on the position and conformal time, $\tau= \int dt/a(t)$, with $t$ the measurable cosmic time. 

The inflaton background field, hereafter denoted by $\phi$ instead of $\phi_0$, for simplicity, speeds the universe up by virtue of a \emph{quasi-de Sitter phase}, yielding  the unperturbed metric tensor, 
\be \label{homet}
g_{\mu \nu}= a^2(\tau) \eta_{\mu \nu},
\ee
where $\eta_{\mu \nu}$ is the Minkowski metric tensor and we assume \cite{boya}
\be \label{quasids}
a(\tau)= -\frac{1}{H_I \left(\tau-2\tau_R\right)^{1+\epsilon}}.
\ee
At this stage, Eq. \eqref{quasids} deserves some additional comments. Particularly, the time $\tau_R$ describes transition to the radiation dominated era, while $H_I$ provides the Hubble parameter during inflation, up to corrections of first order. Moreover, $\epsilon$ represents slight deviations from a purely de Sitter phase. By calculating the slow-roll parameter, we can precisely identify it with the latter.

As stated above, the presence of inflaton fluctuations induces perturbations on the background spacetime, leading to the perturbed metric tensor
\be \label{pertmet}
g_{\mu \nu}= a^2(\tau) \left( \eta_{\mu \nu}+ h_{\mu \nu} \right),\ \ \ \ \lvert h_{\mu \nu} \rvert \ll 1.
\ee
Selecting now the longitudinal, or conformal Newtonian gauge \cite{bau}, it can be shown that scalar perturbations associated with $\phi$ become particularly simple, 
\be \label{pertmatr}
h_{\mu \nu}=\text{diag}\left( 2\Psi, 2\Psi, 2\Psi, 2\Psi \right),
\ee
where the perturbation potential $\Psi$ satisfies \cite{rio}
\be \label{perturb}
\Psi^\prime+\mathcal{H} \Psi= \epsilon \mathcal{H}^2 \frac{\delta \phi}{\phi^\prime},
\ee
with $\mathcal{H}=a^\prime/a$ and the prime denoting derivative with respect to conformal time. 

Before studying spacetime perturbations for some specific models of inflation and computing the corresponding particle production, we briefly discuss the evolution of the spectator field, focusing on

\begin{itemize}
    \item[-] its dynamics during the slow-roll regime,
    \item[-] the subsequent transition of this field to the radiation era.
\end{itemize}

We underline that, by assuming an instantaneous transition to the radiation era, we are neglecting the effects of reheating on GPP. Nevertheless, we will see that super-Hubble modes feature very slow dynamics at the end of inflation, so they are typically unaffected by the microphysical processes of thermalization that should take place after the slow-roll regime \cite{boya}. It is clear, however, that this picture induces an approximation that may overestimate the production itself. A more comprehensive analysis needs to take into account more refined matching conditions and the effects of backreaction, which will be both subject of future investigations. For the moment, we focus on the slow-roll dynamics of the spectator field and the transition occurring between inflation and radiation epochs.


\subsection{Spectator field dynamics during inflation} \label{sec2B}

Following standard approaches \cite{gpp2,boya}, we consider the conformally rescaled spectator field, 
\be \label{resc}
\chi({\bf x}, \tau)= a(\tau) \varphi({\bf x}, \tau),
\ee
quantized by
\be \label{quantspec}
\hat{\chi}({\bf x}, \tau)= \frac{1}{(2\pi)^{3/2}} \int d^3k \left[ \hat{a}_{\bf k} g_k(\tau) e^{-i{\bf k} \cdot {\bf x}} + \hat{a}_{\bf k}^\dagger g_k^*(\tau) e^{i{\bf k}\cdot {\bf x}} \right],
\ee
where we introduce the comoving momentum, $k$, and the field modes, $g_k(\tau)$, satisfying the differential equation,
\be \label{diffmod}
g_k^{\prime \prime} (\tau)+ \left[ k^2+m^2a^2-\frac{a^{\prime \prime}}{a}(1-6\xi) \right] g_k(\tau)=0.
\ee
Defining now
\be \label{case}
g_k(\tau) = \begin{cases}
    g_k^<(\tau)\ \ \ \ \text{for}\ \tau < \tau_R, \\
    g_k^>(\tau)\ \ \ \ \text{for}\ \tau > \tau_R,
\end{cases}
\ee
we recall the ansatz of Eq. \eqref{quasids} to obtain the mode evolution during inflation, namely
\be \label{explmod}
\frac{d^2}{d\eta^2}g_k^<+ \left[ k^2- \frac{1}{\eta^2}\left( (1-6\xi)(2+3\epsilon)-\frac{m^2}{H_I^2} \right) \right] g_k^<=0,
\ee
where $\eta=\tau-2\tau_R$. Notice that we  exploit the fact that $a^{\prime \prime}/a \simeq (2+3\epsilon)/\eta^2$, since $\epsilon \ll 1$ throughout the slow-roll phase. 

The solutions of Eq. \eqref{explmod} can be expressed in the form
\be \label{hanksol}
g_k^<(\eta)= \sqrt{-\eta} \left[ c_1(k) H_{\nu}^{(1)}(-k\eta)+c_2(k) H_{\nu}^{(2)}(-k\eta) \right],
\ee
where $H_{\nu}^{(1)}$ and $H_{\nu}^{(2)}$ are Hankel functions and
\be \label{hankind}
\nu= \sqrt{\frac{1}{4}+(1-6\xi)(2+3\epsilon)-\frac{m^2}{H_I^2}}.
\ee
The integration constants, $c_1(k)$ and $c_2(k)$, are determined by choosing the \emph{in} vacuum state for the field. A common ansatz consists in employing the \emph{Bunch-Davies vacuum state} \cite{bunch1,bunch2,bunch3}, that requires the asymptotic condition,
\be \label{asymp}
g_k(\eta)  \xrightarrow[\eta \rightarrow -\infty]{} \frac{e^{-ik\eta}}{\sqrt{2k}}.
\ee
This choice implies  $c_1(k)=\sqrt{\pi}e^{i(\nu +\frac{1}{2})\frac{\pi}{2}}/2$ and $c_2(k)=0$, so the rescaled field modes take the form
\be \label{influct}
g_k^< (\eta)= \frac{\sqrt{-\pi\eta}}{2} e^{i\left( \nu+ \frac{1}{2}\right) \frac{\pi}{2}}  H^{\left(1\right)}_{\nu}\left(-k\eta\right).
\ee
Exploiting the asymptotic behaviour of Hankel functions, one can show that on super-Hubble scales, $k \ll aH_I$, the original field modes are nearly frozen, while they oscillate on sub-Hubble scales, $k \gg aH_I$ \cite{rio, bel1}.


\subsection{The spectator field transition to radiation era} \label{sec2C}

At  $\tau=\tau_R$, we assume an instantaneous transition from inflation to the radiation dominated phase, whose dynamics is still described by the metric tensor of Eq. \eqref{homet}, with $a(\tau)=H_R \tau$. 

From the continuity of the scale factor at $\tau_R$, we have\footnote{We neglect the dynamics of perturbations at the end of slow-roll, which in principle may affect the matching conditions. However, the presence of backreaction mechanisms at the end of inflation is expected to reduce the net amount of inhomogeneities, see e.g. \cite{backr}.}
\be \label{matchco}
\frac{1}{H_I \left(\tau_R\right)^{1+\epsilon}}= H_R \tau_R,
\ee
that, under the assumption $\epsilon \ll 1$, gives
\be \label{tratime}
\tau_R \simeq \frac{1}{\sqrt{H_I H_R}},
\ee
where $H_R \simeq 10^{-35}$ eV \cite{boya} and $H_I$ can be derived by fixing the energy scales of inflation. 

During the radiation era, from Eq. \eqref{diffmod} we can write
\be \label{radmod}
\frac{d^2}{d\tau^2} g_k^>+\left[ k^2+m^2 H_R^2 \tau^2  \right]g_k^>=0,
\ee
which is solved in terms of parabolic cylinder functions. The general solution of Eq. \eqref{radmod} can be expressed in the form
\be \label{radsol}
g_k^>(\tau) = \alpha_k f_k(\tau)+\beta_kf^*_k(\tau),
\ee
where $\alpha_k$, $\beta_k$ are known as Bogoliubov coefficients. The modes $f_k(\tau)$ satisfy Eq. \eqref{radmod} with asymptotic boundary condition
\be \label{asycond}
f_k(\tau) \xrightarrow[\tau \rightarrow -\infty]{} \frac{e^{-i \int^\tau \omega_k(\tau^\prime) d\tau^\prime }}{\sqrt{2 \omega_k(\tau)}},
\ee
where $\omega_k(\tau)= \sqrt{k^2+m^2H_R^2\tau^2}$. In order to properly define the notion of particle and vacuum state during the radiation phase, the \emph{adiabatic condition} 
\be \label{adcond}
\frac{\omega_k^\prime(\tau)}{\omega_k^2(\tau)} \ll 1.
\ee
should be satisfied \cite{gpp2}. An upper bound to this ratio is given by modes with negligible momentum, for which the adiabatic approximation gives
\be \label{adcond0}
a(\tau) \gg \sqrt{H_R/m}.
\ee
It can be shown that this condition is verified well before matter-radiation equality, even in case of ultralight DM candidates with $m \ll 1$ eV \cite{boya2}.\\

Accordingly, we can properly associate \emph{out} particle states to the modes $f_k(\tau)$, which are normalized via the Wronskian condition
\be \label{wro}
f_k^\prime(\tau) f_k^*(\tau)-f_k(\tau) f_k^{\prime *}(\tau)=-i.\\[2pt]
\ee
Spectator field modes also require proper matching conditions at $\tau=\tau_R$,  to ensure continuity of the field energy density at the transition \cite{boya}. Thus, imposing
\be
\begin{aligned}
&g_k^<(\tau_R) = g_k^>(\tau_R)\,, \label{cond}\\
&\frac{d}{d \tau} g_k^<(\tau) \bigg\lvert_{\tau_R}= \frac{d}{d \tau} g_k^>(\tau) \bigg\lvert_{\tau_R},
\end{aligned}
\ee
and exploiting Eq. \eqref{wro}, one obtains the Bogoliubov coefficients associated with this transition, namely
\begin{align}
    & \alpha_k= i \left[ g_k^{\prime <} (\tau_R) f_k^*(\tau_R) -g_k^< (\tau_R)f_k^{\prime*} (\tau_R) \right], \label{abo}\\
    & \beta_k= -i \left[ g_k^{\prime <} (\tau_R) f_k(\tau_R) -g_k^< (\tau_R)f_k^{\prime} (\tau_R)     \right].
\end{align}
This implies that the field expansion at $\tau > \tau_R$ can be written as
\begin{align} \label{fieldexp2}
\hat{\chi}({\bf x}, \tau)&= \frac{1}{(2\pi)^{3/2}} \int d^3k \left[ \hat{a}_{\bf k} g_k^>(\tau) e^{-i{\bf k} \cdot {\bf x}} + \hat{a}_{\bf k}^\dagger g_k^{*>}(\tau) e^{i{\bf k}\cdot {\bf x}} \right] \notag \\
& = \frac{1}{(2\pi)^{3/2}} \int d^3k \left[ \hat{b}_{\bf k} f_k(\tau) e^{-i{\bf k} \cdot {\bf x}} + \hat{b}_{\bf k}^\dagger f_k^{*}(\tau) e^{i{\bf k}\cdot {\bf x}} \right],
\end{align}
where we  introduced $
\hat{b}_{\bf k}=\alpha_k \hat{a}_{\bf k}+ \beta_k^* \hat{a}^\dagger_{-{\bf k}}$.

\begin{figure}
    \centering
    \includegraphics[scale=0.67]{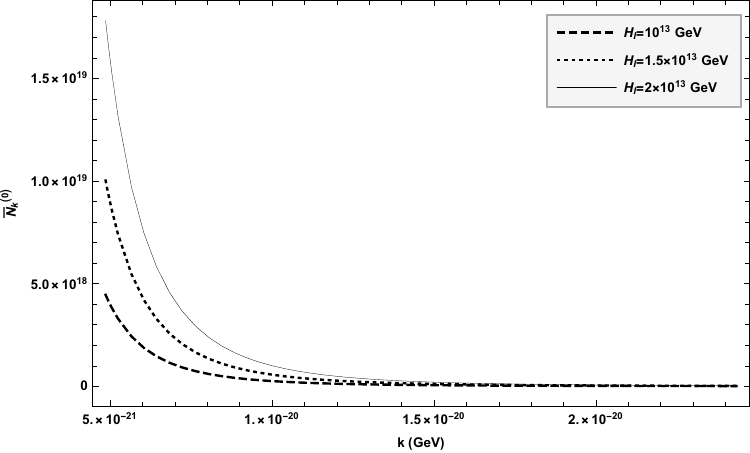}
    \caption{Rescaled number density $\bar{N}_k^{(0)}$ as function of the momentum $k \in \left[ 10^{-5}/\tau_R, 10^{-4}/\tau_R  \right]$, for typical values of the Hubble parameter during inflation. We set $\tau_R \simeq 2.05 \times 10^{15}$ GeV$^{-1}$, $m=10^{-14}$ GeV and $\xi=0$. In case of bosonic fields, GPP is generally more efficient for modes which are super-Hubble at the end of inflation.}
    \label{figzero}
\end{figure}

\subsection{Producing particles from the spectator field}

By virtue of the above results, we can now identify $\hat{b}_{\bf k}, \hat{b}_{\bf k}^\dagger$ as the ladder operators corresponding to \emph{out} particle states, obeying canonical quantization conditions. Since \emph{in} and \emph{out} vacua are different in general, due to the background expansion, a certain amount of particles is produced via the GPP mechanism.

In the Heisenberg picture, the final comoving number density of spectator field particles reads
\be \label{numpart}
N_k^{(0)} \equiv \frac{1}{a^2(\tau_R)} \langle 0 \lvert \hat{b}^\dagger_{\bf k} \hat{b}_{\bf k} \rvert 0 \rangle
= \frac{\lvert \beta_k \rvert^2}{a^2(\tau_R)},
\ee
where $\lvert 0 \rangle$ is the initial Bunch-Davies vacuum state, satisfying the condition,  
$\hat{a}_{\bf k} \lvert 0 \rangle=0$, $\forall\  {\bf k}$.

Thus, Eq. \eqref{numpart} implies that the initial vacuum state of the field is no longer seen as a vacuum in the \emph{out} region. Hence, we can interpret $N_k^{(0)}$ as the number density of particles asymptotically produced from cosmic expansion, i.e., in terms of a gravitational production obtained from vacuum.

Defining now the quantities,
\begin{align} 
& R_k= \frac{2^{3/4}}{\lvert \alpha \rvert^{1/4}} \left \lvert \frac{\Gamma\left( \frac{3}{4}-i\frac{\lvert \alpha \rvert}{2} \right)}{\Gamma\left( \frac{1}{4}-i\frac{\lvert \alpha \rvert}{2} \right)} \right \rvert^{1/2}, \label{gppar3} \\
& \kappa= \sqrt{1+e^{-2\pi \lvert \alpha \rvert}}-e^{-\pi \lvert \alpha \rvert}, \label{gppar2} \\
& \alpha=- \frac{k^2}{2mH_R}, \label{gppar1},
\end{align}
it can be shown that, in the limit of minimal coupling $\xi=0$, the number density of gravitationally produced particles for super-Hubble modes $k \tau_R \ll 1$ reads \cite{boya,boya2}
\be \label{numbmin}
N_k^{(0)}= \frac{1}{4R_k^2 \delta^4\ a^2(\tau_R)}\left[ \kappa \left( \frac{R_k^2 \delta}{2}-1 \right)^2 + \frac{1}{\kappa} \left( \frac{R_k^2 \delta}{2}+1 \right)^2  \right],
\ee
where we introduced the additional parameter, $\delta\equiv k\tau_R$. 

It is quite convenient to quantify the rescaled number density,
\be \label{rescnu}
\bar{N}^{(0)}_k \equiv N_k^{(0)} a^2(\tau_R),
\ee
drawn in Fig. \ref{figzero}, where we explore super-Hubble momenta within the range $k \in \left[ 10^{-5}/\tau_R, 10^{-4}/\tau_R \right]$, by assuming different values for $H_I$.

We remark that the number density is strongly peaked at low momentum, due to the bosonic nature of the spectator field considered. At the same time, GPP is more efficient at larger $H_I$, since in this case there is more energy to be converted into particles. 

In the next section, we will include inflationary perturbations in this framework, showing how the presence of spacetime inhomogeneities is able to enhance the total number of particles produced, also allowing for mode-mixing in particle creation.


\section{Geometric contribution to particle creation} \label{sec3}

Particle production from spacetime perturbations represents an alternative mechanism to the widely-studied GPP approach \cite{fri, ces}. In particular, during inflation the presence of inhomogeneities can be traced back to the quantum fluctuations of the inflaton field, which are the fundamental seeds for structure formation in our universe. 

From Eq. \eqref{pertmet}, the first-order interaction Lagrangian density describing the coupling between perturbations and a given quantum field can be written in the form
\be \label{intlag}
    \mathcal{L}_{I}=-\frac{1}{2}\sqrt{-g_{(0)}}H^{\mu\nu}T^{\left(0\right)}_{\mu\nu},
\ee
where $T_{\mu \nu}^{(0)}$ is the zero-order energy-momentum tensor for the field, $g_{(0)}$ the determinant of the background unperturbed metric tensor and $H_{\mu \nu}= a^2(\tau) h_{\mu \nu}$. When dealing with the spectator scalar field $\varphi$ introduced in Sec. \ref{sec2}, we have \cite{fri}
\begin{align} \label{emtens}
T_{\mu \nu}^{(0)}=& \partial_{\mu} \varphi  \partial_{\nu}\varphi-\frac{1}{2} g_{\mu \nu}^{(0)} \left[ g^{\rho \sigma}_{(0)}\  \partial_{\rho} \varphi\  \partial_{\sigma} \varphi - m^2 \varphi^2  \right] \notag \\ 
&- \xi \left[ \nabla_{\mu} \partial_{\nu}- g_{\mu \nu}^{(0)} \nabla^\rho \nabla_\rho+R_{\mu \nu}^{(0)}-\frac{1}{2} R^{(0)} g_{\mu \nu}^{(0)}   \right] \varphi^2.
\end{align}
Moving now to the interaction picture, it can be shown that the first-order $S$ matrix in Dyson's expansion associated with $\mathcal{L}_I$ reads 
$\hat{S} \simeq 1 + i \hat{T} \int d^4x \mathcal{L}_I$.

Since both the field potential and the field-curvature coupling term are quadratic in $\varphi$, particles are produced in pairs at first perturbative order. We can write the corresponding probability amplitude as \cite{bel1}
\begin{align} \label{compact}
\mathcal{C}_{{\bf k}_1,{\bf k}_2} & \equiv \langle {\bf k}_1,{\bf k}_2 \lvert \hat S \rvert 0 \rangle \notag \\ 
& = -\frac{i}{2\ (2\pi)^3} \int d^4x\  2a^2\big(A_0({\bf x}, \tau)
+A_1({\bf x},\tau) \notag \\
&\ \ \ \ \ \ \ \ \ \ \ \ \ \ \ \ \ \ \ \ \ \ \ \ \ \ +A_2({\bf x},\tau)+A_3({\bf x},\tau) \big),
\end{align}
where
\begin{widetext}
\begin{align} \label{Atime}
A_0({\bf x},\tau)= 2 \Psi \bigg[&\partial_0 \varphi_{k_1}^*\  \partial_0  \varphi_{k_2}^*-\frac{1}{2} \big( \eta^{\rho \sigma} \partial_\rho \varphi_{k_1}^* \ \partial_\sigma \varphi_{k_2}^*-m^2a^2 \varphi_{k_1} \varphi_{k_2} \big) \notag \\
&- \xi \bigg(\partial_0 \partial_0-\frac{a^\prime}{a}\partial_0 -\eta^{\rho \sigma}\partial_\rho \partial_\sigma -3\left( \frac{a^\prime}{a} \bigg)^2   \right) \varphi_{k_1}^* \varphi_{k_2}^* \bigg] e^{-i({\bf k_1}+{\bf k_2})\cdot {\bf x}}\notag\\
\end{align}
and, similarly,
\begin{align} \label{Aspace}
    A_i({\bf x},\tau)=2 \Psi \bigg[&\partial_i  \varphi_{k_1}^*\  \partial_i \varphi_{k_2}^* +\frac{1}{2} \big( \eta^{\rho \sigma} \partial_\rho  \varphi_{k_1}^* \ \partial_\sigma  \varphi_{k_2}^*-m^2a^2 \varphi_{k_1} \varphi_{k_2} \big)\notag \\
    &- \xi \bigg(\partial_i \partial_i+\frac{3a^\prime}{a}\partial_0+ \frac{2a^{\prime \prime}}{a}+\eta^{\rho \sigma}\partial_\rho \partial_\sigma
    -\left( \frac{a^\prime}{a} \right)^2  \bigg)  \varphi_{k_1}^* \varphi_{k_2}^* \bigg] e^{-i({\bf k_1}+{\bf k_2})\cdot {\bf x}},\notag\\
\end{align}
\end{widetext}
for $i=1,2,3$. In Eqs. \eqref{Atime}-\eqref{Aspace}, we reintroduced the original field modes during inflation, 
\be \label{orig}
\varphi_k(\tau)=\frac{g_k^<(\tau) }{a(\tau)},
\ee
in order to properly compute the amount of spectator particles produced.
For each particle pair, the final state can be written in the form,
\be \label{finstat}
\lvert \Psi \rangle = \hat{S} \lvert 0_{{\bf k}_1}; 0_{{\bf k}_2} \rangle = \mathcal{N} \left( \lvert 0_{{\bf k}_1}; 0_{{\bf k}_2} \rangle + \frac{1}{2} \mathcal{C}_{{\bf k}_1,{\bf k}_2} \lvert 1_{{\bf k}_1};1_{{\bf k}_2} \rangle  \right),
\ee
where the normalization factor $\mathcal{N}$ is derived as usual from the condition $\langle \Psi \rvert \Psi \rangle=1$. 

The comoving number density associated with a geometric production of  particles can be then computed at first and second perturbative order, giving respectively
\begin{align}
    &N^{(1)}_k= \lvert \mathcal{N} \rvert^2\  \delta^3 ({\bf k_1}+{\bf k_2})\   {\rm Re} \left[ \mathcal{C}_{{\bf k}_1,{\bf k}_2} (\alpha_{\bf k_1}\beta_{\bf k_1}+\alpha_{\bf k_2}\beta_{\bf k_2})  \right],\label{numpart1}\\
    &N^{(2)}_{k_1,k_2}= \lvert \mathcal{N} \rvert^2\  \lvert \mathcal{C}_{{\bf k}_1,{\bf k}_2}  \rvert^2  \left( 1+ \lvert \beta_{{\bf k}_1} \rvert^2+ \lvert \beta_{{\bf k}_2} \rvert^2 \right).\label{numpart2}
\end{align}


\subsection{Amplitudes and orders of particle production}

It is relevant to stress that probability amplitudes for pair production are typically small in our perturbative approach. We thus have $\lvert \mathcal{N} \rvert^2 \simeq 1$ and we can neglect the normalization constant in the further computations. Analogously, we underline that, when computing number densities in the interaction picture, the zero-order term of Eq. \eqref{numpart} also acquires a normalization factor. Specifically, the final state of the system is modified by the interaction itself, implying that the contribution of Eqs. \eqref{numpart1}-\eqref{numpart2} with respect to the background GPP term is always independent from the normalization procedure. 

In particular, we notice that the first order term in Eq. \eqref{numpart1} involves creation of particles with opposite momenta, thus only increasing the total number of particle-antiparticle pairs. For this reason, we will focus on the second order term, which instead introduces mode-mixing in particle production.  
In particular, we are interested in superhorizon pair production, so we pick one mode on super-Hubble scales and the other on sub-Hubble ones, namely
\be
\begin{aligned} \label{modcond}
    &a(\tau_i)H_I < \lvert  {\bf k}_1 \rvert < a(\tau)H_I, \\
    &a(\tau)H_I < \lvert {\bf k}_2 \rvert < a(\tau) M_{\rm pl},
\end{aligned}
\ee
where $\tau_i$ is the initial time for inflation and the ultraviolet cutoff is given by the Planck mass $M_{\rm pl}$. Moreover, we assume that there are no super-Hubble modes at the beginning of inflation\footnote{This approach has been recently employed to compute the entanglement entropy of cosmological perturbations across the Hubble horizon, see e.g.  \cite{ent3,ent4}.}.

Exploiting the properties of Hankel functions, from Eqs. \eqref{influct} and \eqref{orig}, we can write \cite{bel1}
\begin{align}
    &\varphi_k^{\rm super} \simeq e^{i\left(\nu-\frac{1}{2}\right)\frac{\pi}{2}}2^{\left(\nu-\frac{3}{2}\right)}\frac{\Gamma\left(\nu\right)}{\Gamma\left(\frac{3}{2}\right)}\frac{H_{I}}{\sqrt{2k^{3}}}\left(\frac{k}{aH_{I}}\right)^{\frac{3}{2}-\nu}, \label{supfluc} \\
    \,\nonumber\\
    &\varphi_k^{\rm sub} \simeq \frac{1}{\sqrt{2k}}  \frac{e^{i \left( \nu+ \frac{1}{2} \right) \frac{\pi}{2}}\ e^{i\left( -k\tau-\frac{\pi}{2}\nu-\frac{\pi}{4}  \right)}}{a}. \label{subfluc}
\end{align}
In the following, we specify our calculations to some relevant inflationary potentials, in order to compute the amount of geometric particle produced during the slow-roll phase. 


\section{Theoretical consequences of inflationary particle production}\label{sezione4}

As above stated, our particle computation depends on the underlying inflationary potential. Thus,  to accurately compute the production of geometric particles during the slow-roll phase, it is mandatory to meticulously identify the most promising approaches that agree with current  observations. The Planck satellite's numerical findings suggest that two categories of potentials remain viable, namely large and small field potentials \cite{planck}.

\begin{figure} [ht]
    \centering
    \includegraphics[scale=0.68]{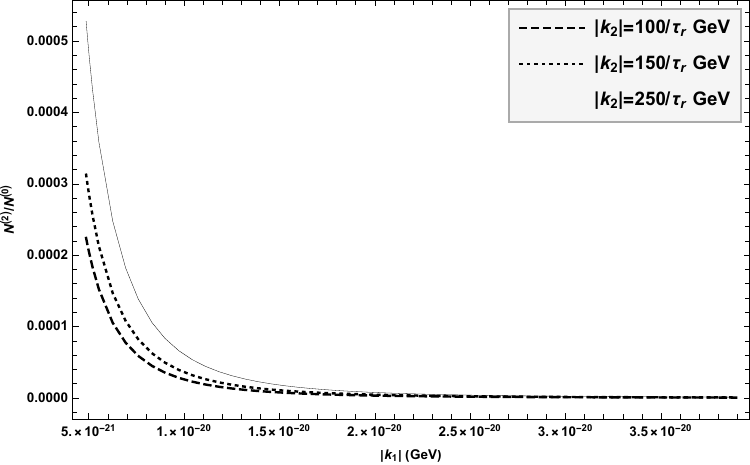}
    \caption{Ratio between the number density for ``geometric" particles $N^{(2)}_{k_1,k_2}$ and the unperturbed density $N^{(0)}_k$, assuming the Starobinsky potential to drive inflation. The ratio is plotted as function of the super-Hubble mode $\lvert {\bf k}_1 \rvert \in  \left[ 10^{-5}/\tau_R, 10^{-4}/\tau_R \right]$. We set $\xi_\varphi=10^{-4}$, $\epsilon=10^{-3}$, $\phi(\tau_i)= 5\ M_{\rm pl}$, $\Lambda^4=10^{64}$ GeV$^4$ and $m=10^{-14}$ GeV.}
    \label{figstar}
\end{figure}

Even though appealing, the class of small field potentials is expected to provide a very small geometric particle production across the Hubble horizon, being incompatible with the possibility that DM can arise from perturbative approaches, see e.g. \cite{bel1, giam}.

In other words, inflationary particle production from inhomogeneities is typically inefficient if the energy in the inflaton field is not large enough. This implies that the substantial energy released during inflation has the potential to be physically converted into particles.

In this respect, we focus on two main large-field inflationary potentials:
\begin{itemize}
    \item[-] The Starobinsky potential \cite{staro1}, that is characterized by the inclusion of the quadratic term $R^2$ in the Hilbert-Einstein action, currently representing the leading candidate to describe inflation. 
    \item[-] The nonminimally coupled fourth order chaotic potential. Here, the fourth order potential alone is unsuitable to describe inflation \cite{planck}, albeit its coupling with $R$  quite evidently  candidates it as a still viable inflationary framework.
\end{itemize}

Below, we discuss the production of particles in both the aforementioned schemes.

\subsection{Particles produced from Starobinsky potential} \label{sec3A}

Here, the metric tensor can be conformally rescaled into the Einstein frame, where the action takes the form of Eq. \eqref{inflag}, with corresponding potential \cite{staro2,staro3}
\be \label{starpot}
V(\phi)= \Lambda^4\left( 1-e^{-\sqrt{2/3} \phi/M_{\rm pl}} \right)^2,
\ee
and $\Lambda^4$ describes the energy scales of inflation. Recalling Eq. \eqref{infans}, the background dynamics of this effective field during slow-roll is given by
\be \label{slowbac}
3 \mathcal{H} \phi^\prime \simeq - V_{, \phi} a^2,
\ee
where we have introduced the compact notation $V_{, \phi} \equiv \partial V/ \partial \phi$ and the scale factor during inflation has been defined in Eq. \eqref{quasids}.
The corresponding fluctuation modes are described by \cite{rio}
\be \label{flures}
\delta \chi_k^{\prime \prime}+ \left[ k^2-\frac{1}{\eta^2} \left( 2+9\epsilon- \frac{ V_{\phi \phi}}{H_I^2} \right)    \right] \delta \chi_k=0,
\ee
where the fluctuation field has been rescaled as usual by $\delta \chi_k= \delta \phi_k a$. This equation admits solutions in terms of Hankel functions, provided the potential term is substituted by its mean value during slow-roll \cite{bel1}.

\begin{figure} [ht!]
    \centering

    \includegraphics[scale=0.68]{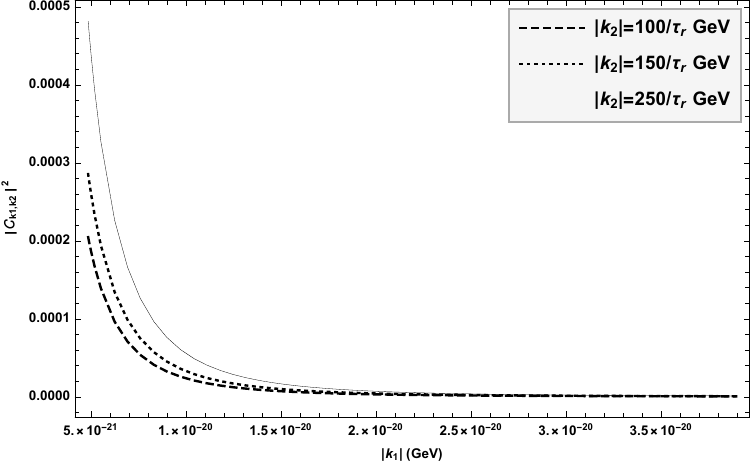}
    \caption{Ratio between the number density for ``geometric" particles $N^{(2)}_{k_1,k_2}$ and the unperturbed density $N^{(0)}_k$, assuming a quartic symmetry-breaking potential. The ratio is plotted as function of the super-Hubble mode $\lvert {\bf k}_1 \rvert\in \left[ 10^{-5}/\tau_R, 10^{-4}/\tau_R \right]$. We set $\xi_\varphi=\xi_\phi=10^{-4}$, $\epsilon=10^{-3}$, $\phi(\tau_i)= 5\ M_{\rm pl}$, $\lambda=2.9\times 10^{-15}$ and $m=10^{-14}$ GeV.}
    \label{figqua}
\end{figure}
Once obtained the background and fluctuation dynamics, the perturbation potential $\Psi$ can be derived from Eq. \eqref{perturb} and inserted in Eq. \eqref{intlag} after proper normalization\footnote{The amplitude of perturbations is typically fixed at horizon crossing $k=a(\tau)H_I$ to a sufficiently small value, namely $\lvert \Phi_{k=aH_I} \rvert \ll 1$, see e.g.  \cite{nohair}.},  to compute the amount of spectator field particles arising from perturbations. In Fig. \ref{figstar}, we show the ratio $N^{(2)}/N^{(0)}$ as function of the super-Hubble mode $k_1$. In particular, the probability amplitude for perturbative production is evaluated in the range $\tau \in \left[ 0, \tau_R\right]$, in order to exploit the simplified expression of Eq. \eqref{supfluc} for the modes under investigation\footnote{As discussed elsewhere \cite{bel1,bel2}, bosonic particle production from inhomogeneities is typically more efficient close to the IR cutoff, in analogy with GPP. For this reason, perturbative production is negligible for spectator field modes that leave the Hubble horizon after $\tau=0$, i.e., modes with $k_1 > 1/\left( 2\tau_R \right)$.}. At the same time, we neglect perturbative production during the radiation era, where the contribution of inhomogeneities is expected to be much smaller due to the presence of other quantum fields and possible backreaction mechanisms.


\subsection{Particles produced from a nonminimal fourth order chaotic potential} \label{sec3B}

As an alternative scenario, we discuss a nonminimally coupled inflaton field driven by a quartic symmetry-breaking potential. This gives
\be \label{quapot}
V(\phi)= \frac{\lambda}{4}\left( \phi^2-v^2 \right)^2+ \frac{1}{2} \xi_{\phi} R\phi^2,
\ee
where $v$ is the vacuum expectation value of the inflaton field and $\lambda$ a self-coupling constant. In case of positive coupling constant, quartic chaotic inflation is not expected to work, unless the inflaton coupling to curvature is sufficiently small \cite{futa,faki} (see also \cite{gum}). This model has been recently considered for perturbative particle production in inflationary scenarios \cite{bel1,efive} and it may also allow to identify the Standard Model Higgs field as the inflaton \cite{higgs1,higgs2,higgs3}.

Following the same steps of Sec. \ref{sec3A}, we can derive the dynamics of the perturbation potential in this model and then compute the corresponding number densities of particles arising from inhomogeneities. In Fig. \ref{figqua} we show again the number density of geometric particles produced at second perturbative order, normalized with respect to the unperturbed GPP contribution.


\section{Consequences and predictions of our scenarios} \label{sezione5}

We here discuss the main implications of our findings in inflationary stages and possible signatures of our scenarios. 

Particularly, let us first observe, from Figs. \ref{figstar}-\ref{figqua}, that perturbative particle production is non-negligible for modes that crossed the Hubble horizon mainly before the cutoff time, $\tau=0$. Second, we also notice that, when approaching the infrared cutoff $\lvert {\bf k}_1 \rvert \simeq a(\tau_i) H_I$, the contribution of inhomogeneities is typically larger. In particular, it can be shown that a perturbative treatment is no longer possible at sufficiently small $\lvert {\bf k}_1 \rvert$, thus requiring a different technique to evaluate the effects of inflaton fluctuations. The above issue is also related to the normalization procedure for the perturbation potential $\Psi$, whose amplitude is typically fixed at horizon crossing \cite{nohair}. Specifically, a correct normalization procedure might depend upon modes in order to guarantee that all the perturbation magnitudes appear the same at the horizon crossing. Likely, this would imply to reformulate  the correct vacuum, modifying the Bunch-Davies choice with a more refined approach. 

Further, we notice that the contribution of inhomogeneities is typically enhanced in case of larger field-curvature coupling constants. In the limit of conformal coupling, $\xi=1/6$, GPP results in negligible densities \cite{boya}, implying that geometric production would become the dominant mechanism for primordial particle creation. A similar result was obtained for the gravitational production of massless fermions during preheating \cite{tsuji}, showing that metric perturbations may have also played an important role at the end of inflation. 

To summarize, particle production arising from inhomogeneities can significantly affect the total number density of spectator field particles created up to the radiation era. For this reason, if DM has been produced via purely gravitational mechanisms, the presence of inhomogeneities should be taken into account when computing the corresponding particle abundance.

Last but not least, we also remarked how spacetime inhomogeneities are responsible for mode-mixing in particle production, that is not conversely found in unperturbed GPP scenarios, where only particle-antiparticle pairs can be generated, i.e., with opposite momenta. More specifically, during inflation the Hubble horizon emerges as a natural separation scale for modes and superhorizon particle production has been recently investigated for inflaton fluctuations, showing that quantum entanglement can be generated in this process \cite{bel1}. 

Remarkably, since we focused on particle production across the Hubble horizon, we conclude that plausible detectable quantum ``signatures" at late times can occur. Indeed, DM is expected to weakly interact with Standard Model fields, so that some entanglement entropy associated with particle production may have survived after the inflationary epoch. Hence, we underline that the role of entanglement could help to understand the quantum properties of produced particles, thus opening new avenues in the search for DM.


\section{Final outlooks and perspectives}\label{sezione6}

In this work, we investigated the particle production associated with a spectator scalar field, i.e., subdominant with respect to the inflaton, during and after the slow-roll regime. To show how particle production is influenced by the universe expansion, we pictured an instantaneous transition from inflation to the radiation dominated era, neglecting the effects due to reheating in GPP regimes.

In particular, we focused on the contribution associated with inhomogeneous particle production across the Hubble horizon, that can be traced back to the fluctuations of the inflaton field during slow-roll. We thus showed that the number density of particles arising from perturbations is typically non-negligible with respect to the widely-studied quantum GPP contribution, obtained from the unperturbed universe expansion.

We reobtained this outcome in the realms of large-field inflation and, particularly, we focused on two among the most consolidate paradigms describing the inflationary speed up. Specifically, we worked out the Starobinsky and the fourth order nonminimally coupled potentials. The latter represents the most viable large-field model of inflation, conformally equivalent to an extended theory of gravity, whereas the second is a suitable example of chaotic inflation, overcoming the Planck satellite observational constraints. 

We discussed the physical results obtained and, particularly, we showed that the amount of particles obtained is similar in both the aforementioned scenarios. We also argued that geometric particle production across the horizon is expected to be negligible in small-field approaches, since in that case the energy in the inflaton field is significantly smaller throughout the slow-roll regime.

In addition, we observed that the presence of inhomogeneities allows for mode-mixing in particle production, that instead is not found in unperturbed GPP processes, where the total momentum of created particles is necessarily conserved. The presence of mode-mixing may lead to entanglement generation across the Hubble horizon, and we argued that such quantum correlations could have survived after the inflationary epoch due to the weakly interacting nature of DM. Possibility of detecting such particles through entanglement are also discussed above. 

At the same time, we noticed that a perturbative approach to inhomogeneous particle production is not always possible, since the magnitude of inflaton fluctuations becomes typically large on super-Hubble scales. We also pointed out that a more refined approach is needed for the normalization of the perturbation potential, in order to obtain correct amplitudes at horizon crossing for all the modes involved.

As perspectives, further steps would include the effects of reheating in geometric production. Although such effects may be negligible at a first sight, they may affect the total amount of produced particles via the dynamics of preheating metric perturbations, especially for sub-Hubble modes. 

In addition, we intend to study the possible backreaction effects associated with the dynamics of perturbations at the end of inflation and shed further light on a more general non-perturbative approach to inhomogeneous particle production. 

Finally, we plan to extend our treatment to higher spin spectator fields, starting from fermionic ones, with the aim of evaluating other possible DM candidates.

\section*{Acknowledgements}
The work of OL is  partially financed by the Ministry of Education and Science of the Republic of Kazakhstan, Grant: IRN AP19680128.


\end{document}